\documentclass[a4paper,oneside,english]{aa}
\usepackage{amsmath}
\usepackage{graphics}
\usepackage{graphicx}
\usepackage{amssymb}
\usepackage{natbib}
\usepackage{babel}
\usepackage{xspace}
\bibpunct{(}{)}{;}{a}{}{,}
\newcommand{\ascc}   {\mbox{ASCC-2.5}\xspace}

\newcommand{\olin}[1]{\overline{#1}}

\begin{document}

\title{Shape parameters of Galactic open clusters\thanks{The determined
shape parameters for 650 clusters are listed in a table that is available in
electronic form at the CDS via anonymous ftp to
cdsarc.u-strasbg.fr (130.79.128.5)
or via http://cdsweb.u-strasbg.fr/cgi-bin/qcat?J/A+A/}}

\author{N.V.~Kharchenko \inst{1,2,3} \and
        P.~Berczik \inst{1,2} \and
        M.I.~Petrov \inst{2,4} \and
        A.E.~Piskunov \inst{1,3,5} \and
        S.~R\"{o}ser \inst{1} \and
        E.~Schilbach \inst{1} \and
        R.-D.~Scholz \inst{3} }

\offprints{R.-D.~Scholz}

\institute{
Astronomisches Rechen-Institut, Zentrum f\"{u}r Astronomie der Universit\"{a}t
Heidelberg (ZAH), M\"{o}nchhofstra\ss{}e 12-14, D--69120 Heidelberg, Germany\\
email: nkhar@ari.uni-heidelberg.de,
berczik@ari.uni-heidelberg.de,
apiskunov@ari.uni-heidelberg.de,
roeser@ari.uni-heidelberg.de, elena@ari.uni-heidelberg.de
\and
Main Astronomical Observatory, 27 Academica Zabolotnogo Str., 03680
Kiev, Ukraine\\
email: nkhar@mao.kiev.ua, berczik@mao.kiev.ua, petrov@mao.kiev.ua
\and
Astrophysikalisches Institut Potsdam, An der Sternwarte 16,
D--14482 Potsdam, Germany\\
email: nkharchenko@aip.de, apiskunov@aip.de, rdscholz@aip.de
\and
Institut f\"{u}r Astronomie der Universit\"{a}t Wien, T\"{u}rkenschanzstra\ss{}e
17, A-1180 Wien, Austria\\
email: petrov@astro.univie.ac.at
\and
Institute of Astronomy of the Russian Acad. Sci., 48 Pyatnitskaya
Str., 109017 Moscow, Russia\\
email: piskunov@inasan.rssi.ru
}

\date{Received 17 June 2008; accepted 25 November 2008}

\abstract{There are only a few tens of open clusters for which
ellipticities have been determined in the past.}
{In this paper we derive observed and modelled shape parameters
(apparent ellipticity and orientation of the ellipse) of 650
Galactic open clusters identified in the \ascc catalogue.
}
{We provide the observed shape parameters of Galactic open clusters, computed with
the help of a multi-component analysis. For the vast majority of clusters these
parameters are determined for the first time. High resolution
(``star by star'') N-body simulations are carried out with the specially
developed $\phi$GRAPE code providing models of clusters of different initial
masses, Galactocentric distances and rotation velocities.}
{The comparison of models and observations
of about 150 clusters reveals ellipticities of observed clusters which are
too low (0.2 vs. 0.3), and offers the basis to find the main reason for this discrepancy.
The models predict that  after $\approx 50$ Myr clusters reach an  oblate
shape with an axes ratio of $1.65:1.35:1$, and with the major axis tilted by
an angle of  $q_{XY} \approx 30^\circ$ with respect to the Galactocentric
radius due to differential rotation of the Galaxy.
}
{Unbiased estimates of cluster shape parameters requires reliable membership
determination in large cluster areas up to 2-3 tidal radii where the density
of cluster stars is considerably lower than the background. Although
dynamically bound stars outside the tidal radius contribute insignificantly
to the cluster mass, their distribution is essential for a correct
determination of cluster shape parameters. In contrast, a restricted mass
range of cluster stars does not play such a dramatic role, though deep
surveys allow to identify more cluster members and, therefore, to increase
the accuracy of the observed shape parameters.}


\maketitle

\section{Introduction}

Our current project aims at studying the properties of the local population
of Galactic open clusters. The sample contains 650 open clusters and
cluster-like associations identified in the all-sky compiled catalogue of 2.5
million stars \ascc \citep{kha01}. For each cluster, a combined
spatio-kinematic-photometric membership analysis was performed \citep[]{starcat} and
a homogeneous set of different cluster parameters was derived
\citep[][]{clucat,newclu}. In two recent papers \citep[][]{clumart,clumart1}
tidal radii and masses of open cluster were determined and their relation to
cluster ellipticity was briefly discussed. In the present study we discuss in
detail all the issues related to the shape of local clusters.

The shape of a star cluster, as well as its size and mass are the most
important dynamical parameters. They are predesignated already at the stage
of the cluster formation, when the cluster keeps the memory of the size and
shape of the parent cloud. In addition to internal processes
(self-gravitation and rotation) a considerable role in cluster shaping is
played by external forces, e.g. by the Galactic tidal field, which is acting
in two ways \citep[cf.][]{wiel74,wiel85}: a) stretching the cluster into an
ellipsoid directed towards the Galactic centre, and b) producing cluster
tails outpouring from the ellipsoid endpoints (cluster Lagrangian points).
The tails are composed of stars lost by the cluster, which lead and/or trail
the cluster along its orbit due to differential rotation of the Galactic disk
\citep[see e.g.][]{chura06a}. The encounters of star clusters with giant
molecular clouds randomize the regular effect of the Galactic field
\citep[][]{giea06}.  Additionally, molecular clouds produce a screening
effect, when they  partly overlap the clusters.

Most of the above reasons lead to the violation of cluster's sphericity
with different effects on the cluster core and corona. 
The analysis of such violations and their comparison with
predictions of N-body models can shed light on the dynamical history of open
clusters. We assume in the following that open clusters can be represented by
triaxial ellipsoids.

The violations of the sphericity of globular clusters are evident and their
ellipticities were determined for the first time already at the beginning of
the last century. For example, \citet{sheso27} published the estimates of
ellipticity of 75 globular clusters. Currently, shape parameters are
determined for one hundred Galactic globular clusters \citep[]{white87}, i.e.
according to Harris' on-line catalogue\footnote{
http://www.physics.mcmaster.ca/resources/globular.html} for about two thirds
of the known objects in the Galaxy.

The current status of the shape measurements for open clusters is much
poorer. Until now, indications of a flattening were obtained for a few tens
of open clusters only. \citet{ramer98a, ramer98b} and \citet{adea01,adea02}
found evidence of a flattening in the Pleiades and in Praesepe. Their results
support the predictions of Wielen's model that explains the ellipsoidal form
of clusters by tidal coupling with the Galaxy. In contrary, the flattening
found in the Hyades \citep{oo79} deviates from theoretical expectations.
Recently, \citet{chen04} published morphology parameters, including cluster
ellipticities, for 31 Galactic open clusters, located preferentially in the
anticentre direction of the Galaxy.

With this paper we fill this gap in the parameter list of open clusters and
determine the shape parameters for all 650 open clusters in our sample
\citep{clucat,newclu}. For a comparison with these observed shape parameters
we carried  out a set of high resolution (``star by star'') N-body dynamical
calculations of cluster models with different initial angular momenta located
at different Galactocentric distances in the Milky Way.

The paper has the following structure. In Sec.~\ref{sec_bas-eq} we provide
the basic equations used in the present work; in Sec.~\ref{sec_num-mod} we
describe the N-body models; in Sec.~\ref{obs_mod-res} we compare the observed
shape parameters with model calculations and with literature data; in
Sec.~\ref{sec_concl} we summarize the results.

\section{Calculation of observed parameters of cluster shape}\label{sec_bas-eq}

In this study we use the following coordinate systems: spherical Galactic
coordinates $(l, b)$; the rectangular Galactocentric system $(X, Y, Z)$, with
origin in the Galactic Center, and axes directed to the Sun - $X$, along the
Galactic rotation at the Sun's location - $Y$, and to the North Galactic Pole
- $Z$; and a proper rectangular coordinate system $(X', Y', Z')$ with origin
in the centre of a cluster under consideration. The $X'$-axis is directed
along the projection of the cluster Galactocentric vector onto the Galactic
plane, $Y'$ points to the direction of Galactic rotation and $Z'$ is directed
to the North Galactic  Pole.

Currently, only the two nearest clusters (Ursa Majoris and the Hyades)
provide sufficiently accurate individual distances of their stars allowing
for a direct determination of the 3D structure parameters of these clusters.
For the other clusters we can only consider the stellar distributions in
projection onto the celestial sphere.

To derive the shape parameters of clusters: ellipticity, lengths and
directions of the ellipse axes, we applied a multicomponent analysis of the
positions of cluster members in the standard (or tangential) coordinate
system $x, y$ in each sky area containing a cluster. For each cluster member,
$x, y$ are computed from its Galactic coordinates $l, b$ as:
\[ x =
\frac{\sin (l - l_c)} {\tan b \cdot \sin b_c + \cos b_c \cdot \cos (l -
l_c)}, \] \[ y = \frac{\cos b_c \cdot \tan b - \sin b_c \cdot \cos(l - l_c)}
{\sin b_c \cdot \tan b + \cos b_c \cdot \cos(l - l_c)},
\]
where $l_c, b_c$ are the Galactic coordinates of the cluster centre. The axis
$x$ is parallel to the Galactic plane, and the axis $y$ is pointing to the
North Galactic Pole, positive directions of $x$, $y$ coincide with positive
directions of $l$, $b$, and for the cluster centre $x_c = 0$, $y_c = 0$. The
2nd order momenta of standard coordinates
\[
M_{xx} = \frac{\sum_i x_i^2}{N},\,\,\,   M_{yy} = \frac{\sum_i y_i^2}{N},\,\,\,
M_{xy} = \frac{\sum_i x_i y_i}{N},
\]
where $i = 1, 2,... N$, and $N$ is number of cluster members, are the basis for the
characteristic equation:
\[
\left| \begin{array}{cc} M_{xx}- \Lambda & M_{xy}\\
                        M_{xy} &  M_{yy}- \Lambda \end{array}\right| = 0.
\]
The roots of this equation: $A=\sqrt {\Lambda_1},\,B= \sqrt {\Lambda_2}$ scale
the principal semi-major and semi-minor axes of an apparent ellipse given by
the distribution of cluster members over the sky area. An apparent
ellipticity $e$ (ellipticity hereafter) is then computed as
\begin{equation}
e = 1 - \frac{B}{A}.
\label{e_eqn}
\end{equation}
The orientation of the ellipse is defined by the parameter $q$, which is the
angle between the  Galactic plane and the largest principal axis of the
ellipse:
\begin{equation}
\tan (2q) = \left | \frac{2 \cdot M_{xy}}{M_{xx} - M_{yy}} \right |\,,
\label{q_eqn}
\end{equation}
and varies between $0^\circ$ and $90^\circ$.

The $rms$ errors of these values are computed from:
\[
\varepsilon_e^2 = \frac{A^2 \cdot \varepsilon^2_B + B^2 \cdot \varepsilon^2_A}
{A^4},
\]
\[
\varepsilon_q^2 = \frac{M_{xy}^2 \cdot (\varepsilon^2_{M_{xx}} +
                                       \varepsilon^2_{M_{yy}}) +
                 ( M_{xx}- M_{yy})^2 \cdot \varepsilon^2_{M_{xy}}}
{( M_{xx}- M_{yy})^2 + 4\cdot M_{xy}^2},
\]

where
\[
\varepsilon_{M_{xx}}^2 =  \frac{2 \cdot M_{xx}^2}{N-1},\,\,\,\,
\varepsilon_{M_{yy}}^2 =  \frac{2 \cdot M_{yy}^2}{N-1},
\]
\[
\varepsilon_{M_{xy}}^2 =
  \frac{(M_{xx} \cdot M_{yy} - M_{xy}^2) \cdot (N-1)}{(N-2) \cdot (N-1)},
\]
and $\varepsilon_A$, $\varepsilon_B$ (as well as $\varepsilon_e$ and
$\varepsilon_q$) are calculated applying  the law of error propagation.

\begin{figure}[]
   \centering
\resizebox{\hsize}{!}{\includegraphics[angle=270,clip]{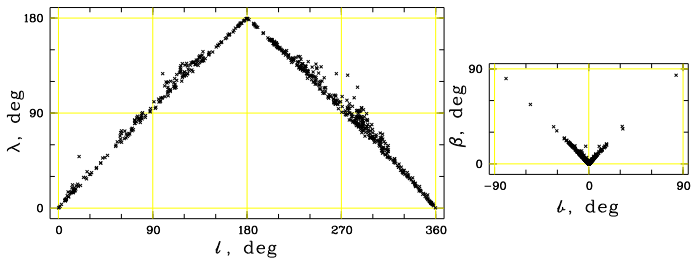}}
\caption{Aspect angles $\lambda$ and $\beta$ versus Galactic coordinates
$l$ and $b$, respectively, for all 650 clusters.
}
\label{proe_fig}
\end{figure}

If the orientations of cluster ellipses are not random, one expects
observing a longitudinal dependence of $e$. Since clusters reside at
different heliocentric distances $d$, it would be more convenient to
consider  a related angle $\lambda$ instead of $l$, where $\lambda$ is the
angle between the projections of the heliocentric and Galactocentric vectors
of a cluster onto the Galactic plane, and can be determined from:
\[
\cos (180^\circ - \lambda) = \frac{d - R_{\odot} \cdot \cos l_c}
{\sqrt{R^2_{\odot} + d^2 - 2 \cdot R_{\odot} \cdot d \cdot \cos l_c}}\,.
\]
The angle $\lambda$  varies between $0^\circ$ and $180^\circ$. Similarly,
$\beta$ is the angle between the projections of the two vectors onto the
meridional plane of a cluster:
\[
\cos (180^\circ - \beta) = \frac{d - R_{\odot} \cdot \cos b_c}
   {\sqrt{R^2_{\odot} + d^2 - 2 \cdot R_{\odot} \cdot d \cdot \cos b_c}}\,,
\]
the angle $\beta$ varies between $0^\circ$ and $90^\circ$. Here $d$ is the
distance of the cluster from the Sun, $R_{\odot} =$ 8.5 kpc is the distance
of the Sun from the Galactic centre. We define $\lambda$ and $\beta$ to be
$\lambda = 180^{\circ}$ at $l = 180^{\circ}$ and $\beta = 0^{\circ}$ at $b =
0^{\circ}$. In Fig.~\ref{proe_fig} we show the values of $\lambda$ and
$\beta$ for all 650 clusters.

Using the equations above, one can compute shape parameters (principal
semi-axes $A$ and $B$, ellipticity $e$ and angles $q$, $\beta$), both for
real and modelled clusters.

\section{Numerical modeling}
\label{sec_num-mod}

A non-sphericity of star clusters is predicted both by theory and numerical
simulations. \citet{wiel74,wiel85} predicted that the ratios of the three
orthogonal axes of the cluster ellipsoid should show the ratio
$(a:b:c)=(2.0:1.4:1.0)$. The largest axis is pointing to the Galactic centre
and the smallest one is directed to the North Pole. These results were
obtained with N-body calculations of 500 particles, and were confirmed later
with somewhat more populated models including up to 1000 particles
\citep[]{terle87}, and up to 2500 particles \citep[]{chura06a}. Such a number
of particles  corresponds to a  cluster initial mass below  $10^3 M_{\odot}$.
Presently there are evidences, however, that average masses of forming star
clusters are larger. For example, \citet[]{fuma} have found, that the average
mass of the Galactic open clusters at birth is equal to $4.5\cdot
10^3\,m_\odot$. Therefore, it is necessary to consider more populated models
than it has been done before.

\subsection{$\phi$GRAPE N-body code}

For our high resolution N-body simulations we use the specially developed
$\phi$GRAPE code. The code itself and also the GRAPE hardware we used
are described in the paper by \citet{Hetal2007} in more detail. Here we just
briefly mention a few of the special features of our code. The program was
already thoroughly tested with different N-body applications, including the
high resolution, direct study of the dynamical evolution of the Galactic
centre with Binary (or Single) Black Holes \citep[]{BMS2005, BMSB2006,
MBL2007}
\footnote{The present version of the code and the full snapshot datasets
analyzed in the paper are publicly available from:
{\tt ftp://ftp.ari.uni-heidelberg.de/pub/staff/ \\
berczik/phi-GRAPE-cluster/}.}\label{models_ref}.

The program acronym $\phi$GRAPE means: P{\it arallel} H{\it
ermite} I{\it ntegration} with GRAPE. The serial and parallel
versions of the program are written from scratch in  ANSI-C and use the
standard MPI library for communication. For the integration of the star
cluster's dynamical evolution inside the Galactic potential we use the
parallel GRAPE systems developed at ARI Heidelberg (GRACE - year 2005) and at
MAO Kiev (GRAPE/GRID - year 2007).

\begin{table}[]
\caption{The parameters of the Galactic potential components.}
\label{gal-par_tbl}
\begin{center}
\begin{tabular}{lcrr}
\hline
\noalign{\smallskip}
 Mass component & $M / m_\odot$ & $a$, kpc & $b$, kpc \\
\noalign{\smallskip}
\hline
\noalign{\smallskip}
 Bulge & $1.4 \cdot 10^{10}$ & 0.0 &  0.3 \\
 Disk  & $9.0 \cdot 10^{10}$ & 3.3 &  0.3 \\
 Halo  & $7.0 \cdot 10^{11}$ & 0.0 & 25.0 \\
\hline
\end{tabular}
\end{center}
\end{table}

The code uses the 4-th order Hermite integration scheme for the particles
with the hierarchical individual block timesteps, together with the parallel
usage of GRAPE6a cards for the hardware calculation of the acceleration
$\vec{a}$ and the first time derivative of the acceleration $\vec{\dot{a}}$
(this term is usually called ``jerk'' in the N-body community).

We specially check the effect of N-body softening, which suppresses
the hard binary formation in our code, on the evolution of the
cluster model. We vary the softening parameter from a few
hundreds down to a few astronomical units and do not find a
significant difference in the evolution of the cluster mass and shape.
We also consider the role of a possible initial mass
segregation and do not observe any dependence of the evolution of the
semimajor axis on this value.

Compared with the previous public version we add two major changes to the
code. First of all we add the possibility to have some external potential,
acceleration and also ``jerk''. For the external potential we choose the form
proposed by \citet{MN1975}. Using such a multi-mass component potential we
can easily approximate the Galaxy's external force, acting on our star
cluster in the Galactic disk at different Galactocentric distances. The
second change includes the possibility to turn on the mass loss due to the
stellar evolution for every modelled star particle. For metallicity-dependent
stellar lifetimes we use the approximation formula proposed by
\citet{RVN1996}: \[ \log \tau = a_0(Z) - a_1(Z) \cdot \log m + a_2(Z) \cdot
\log^2 m, \label{tau} \] where $\tau$ is expressed in years, the stellar mass
$m$ in solar masses, and where $Z$ is the abundance of heavy elements. The
coefficients are defined as: \[ \begin{array}{lllll} a_0(Z)= 10.130 + 0.0755
\cdot \log Z - 0.0081 \cdot \log^2 Z, \\ a_1(Z)= 4.4240 + 0.7939 \cdot \log Z
+ 0.1187 \cdot \log^2 Z, \\ a_2(Z)= 1.2620 + 0.3385 \cdot \log Z + 0.0542
\cdot \log^2 Z. \\ \end{array} \] The mass lost by stars during their
evolution is calculated with the help of tables from the paper of
\citet{vdHG1997} and is approximated analytically via the
metallicity-dependent formula: \[ m_{eject}= -0.4205 \cdot Z^{-0.0177} +
(0.9015 + 0.6294 \cdot Z) \cdot m_{init}. \] For simplicity, we assume in our
model that stars loose their masses permanently with a constant rate:
$-m_{eject}/\tau$.

Since stars lose the bulk of their masses at the end of their
evolution,  we test, also, an alternative mass loss scenario, where we reduce
the  stellar masses just in one timestep at the end of the star life. We find
that the models show almost the same evolutionary patterns for shape parameters
and mass loss  of clusters independent of the version of stellar mass loss
used.

\subsection{Initial conditions for the star cluster}

\begin{table}[t]
\caption{The list of model parameters:  initial mass $M_c(0)$,
number of particles $N$, cluster radius $R_c$, distance from the Galactic
center $R_0$, circular velocity $V_0$, time step $\delta t$, dimensionless
angular velocity $\omega_0$; concentration parameter $W_0 = 6.0$.}
\label{model-list_tbl}
\begin{tabular}{ccccccc}
\hline\noalign{\smallskip}
$M_c(0),$&$N$&$R_c$,&$R_0$,&$V_0$,&$\delta t$,&$\omega_0$\\
$m_\odot$&&    pc&   kpc& km/s& Myr&\\
\noalign{\smallskip}
\hline
\noalign{\smallskip}
 $10^3$      &  4040 &  3 &  7.0 & 236 &2.45 & 0.0, 0.3, 0.6  \\
 $10^3$      &  4040 &  3 &  8.5 & 233 &2.45 & 0.0, 0.3, 0.6  \\
 $10^3$      &  4040 &  3 & 10.0 & 231 &2.45 & 0.0, 0.3, 0.6  \\
 $5\cdot10^3$& 20202 &  7 &  7.0 & 236 &3.91 & 0.0, 0.3, 0.6  \\
 $5\cdot10^3$& 20202 &  7 &  8.5 & 233 &3.91 & 0.0, 0.3, 0.6  \\
 $5\cdot10^3$& 20202 &  7 & 10.0 & 231 &3.91 & 0.0, 0.3, 0.6  \\
 $10^4$      & 40404 & 10 &  7.0 & 236 &4.71 & 0.0, 0.3, 0.6  \\
 $10^4$      & 40404 & 10 &  8.5 & 233 &4.71 & 0.0, 0.3, 0.6  \\
 $10^4$      & 40404 & 10 & 10.0 & 231 &4.71 & 0.0, 0.3, 0.6  \\
\hline
\end{tabular}
\end{table}

For the generation of the initial position and velocity distributions of
cluster particles we use the standard King model. The code for the generation
of the initial rotating star cluster is described in detail by
\citet{ES1999}. We also use their notation to parametrize our rotating King
model family with the concentration parameter $W_0$ and the dimensionless
angular velocity $\omega_0$. Thus each model can be parametrized by the
combination of these two numbers. The larger $W_0$ the larger is the central
concentration. The larger $\omega_0$ the larger is the angular momentum of
the model.

For the present simulations we use the variation with three basic King models
only. We vary only the rotation parameter $\omega_0 = (0.0,~0.3,~0.6)$.
According to Table 1 in \citet{ES1999} the ratio of the ``pure'' rotational
energy to the total kinetic energy of the model for these three cases is
$E_{rot}/E_{kin} = (0.0,~0.105,~0.278)$. The concentration parameter was
always set to $W_0 = 6.0$ (which corresponds to the models with medium
concentration).

As a next step, taking the \citet{salp55} initial mass function (IMF)
$f(m)=dN/dm$, we create the random particle mass distribution:
\[
dn(m)  =  f(m)\cdot dm = {\it C} \cdot m^{-(1+\alpha)} \cdot dm,
\]
with the lower and upper mass limits $m_{l} = 0.08~m_\odot$, $m_{u} =
8.0~m_\odot$ and with Salpeter's slope $\alpha = 1.35$. We use the upper mass
limit of 8.0 $m_\odot$, because we consider a ``pure/classical'' open cluster
(older than 30...40 Myr), when all high mass stars have already finished
their life as a SNII, and swept out all the residual gas left from the
cluster formation process.

The necessary number of particles for a given initial mass of the cluster
$M_c(0)$, can be computed from the assumed IMF as
\[
N = M_c(0) \cdot \frac{-\alpha+1}{-\alpha} \cdot
  \frac{(m_u^{-\alpha} - m_l^{-\alpha})}{(m_u^{-\alpha+1} - m_l^{-\alpha+1})}.
\]
With the adopted IMF parameters for the model cluster and initial masses
$M_c(0) = (10^4, 5\cdot 10^3, 10^3)~m_\odot$ we have the following particle
numbers $N$ = (40404, 20202, 4040).

After our setting of initial masses, positions and velocities of every
particle in the model, we do the standard N-body normalization
\citep{AHW1974} of the model, and rescale the initial cluster data to put our
system into virial equilibrium $E_{gra} = -2 \cdot E_{kin} = 2 \cdot E_{tot}$
with the following  parameters: $G = M_c = 1$ and $E_{tot} = -0.25$. This
step does not change any physical quantities of the modelled star cluster,
but is just very useful for numerical reasons. For a King model with $W_0 =
6.0$ such a normalization produces a half-mass radius of about 0.8 (in
dimensionless units).

After the construction of dimensionless parameters of our model we can easily
extend them to a different physical cluster mass and half-mass radius. In
principle, these two parameters can be set independently, but in order to
reduce the number of independent initial  parameters we decide to use some
physically motivated relation between initial mass and initial half-mass
radius. For this purpose we use the extension of the well known mass vs.
radius relation for molecular clouds and clumps in the Galaxy
\citep[see][]{TH1993, IK2000}, which can be scaled as:
\[
R_c \approx 100
\cdot \sqrt{\frac {M_c} {10^6 ~m_\odot}}~~{\rm pc}
\]
For the three physical
masses chosen we set the corresponding radii to $R_c$ = (10, 7, 3) pc.

In order to check the dependence of the computed evolution of cluster
shape on the adopted mass vs. radius relation, we run a series of models where
we set the normalising factor to 100, 71, and 43 pc. We find that for typical
model parameters, the evolution of the cluster shape does not depend on this
factor, in practice.

\subsection{Galactic rotation curve}

\begin{figure}[]
   \centering
\resizebox{\hsize}{!}{\includegraphics[angle=270,clip]{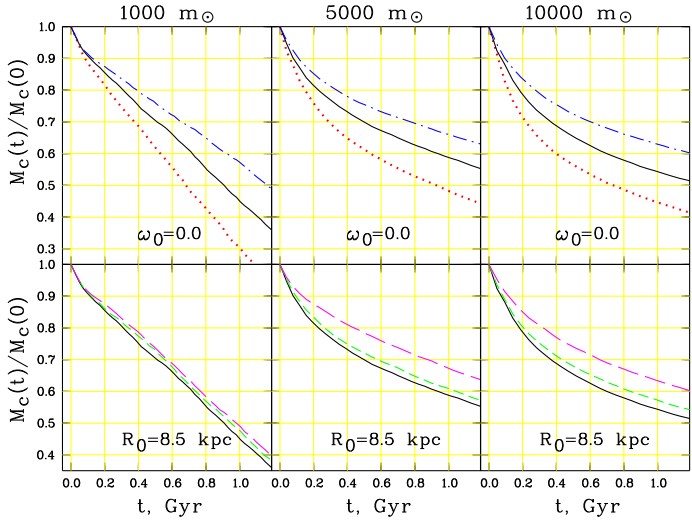}}
\caption{The evolution of the mass of dynamically bound particles for three
cases of the initial mass of the cluster models (three columns  with masses
shown at the top of each column).
The upper panels are constructed for non-rotating models
($\omega_0$ = 0.0) and show the effect of the Galactocentric distance.  The
dotted (red), solid (black) and dashed-dotted (blue) curves are  models at
$R_0 =$ 7, 8.5 and 10 kpc. The bottom panels illustrate the effect of
cluster  rotation at the Solar Galactocentric distance.  The solid (black),
dashed (green) and long dashed (magenta) curves  show models with
$\omega_0 =$ 0.0, 0.3 and 0.6.
}
\label{mass-din_fig}
\end{figure}

For the external Galactic potential shaping a cluster we choose the combined
``Plummer-Kuzmin disk'' form \citep[see][]{MN1975}:
\[
\Phi(r,z) = - \frac{ G \cdot M }{ \sqrt{r^2 + (a + \sqrt{b^2 + z^2} )^2} }.
\]
Coupling such a potential with a three-component Galactic mass distribution
model comprised of ``Bulge'', ``Disk'', ``Halo'' components \citep{DC1995}
one can easily reproduce the observed rotation curve of the Galaxy
\[
\frac{V^2(r)}{r} =
-\nabla\Phi_{Bulge}(r,z) -\nabla\Phi_{Disk}(r,z) -\nabla\Phi_{Halo}(r,z).
\]
In order to bring this into agreement with the kinematics of the open cluster
subsystem in the Solar neighborhood we slightly modify the \citet{DC1995}
parameters. The parameter values consistent with the observed Oort constants
$(A, B) = (14.5 \pm 0.8, -13.0 \pm 1.1)$ km/s/kpc derived by \citet{clupop}
are shown in Table.~\ref{gal-par_tbl}. The adopted values of Oort constants
correspond to  $V/R_\odot = 27.5 \pm 1.3$ km/s/kpc, which at $R_\odot$ = 8.5
kpc gives $V \approx$ 233 km/s. Using these data we derive circular
velocities at distances of 7.5 kpc and 10.0 kpc from the Galactic centre,
too.

\subsection{Model list and model cluster memberships}

As a starting point for our model cluster, we select the position inside the
Galactic disk ($R_0$, 0.0, 0.0) with the corresponding circular velocity
(0.0, $-V_0$, 0.0). For the set of our runs we use three values for the
positions inside the Galactic disk: $R_0$ = (7.0, 8.5, 10.0) kpc with
corresponding circular velocities $V_0$ = (236, 233, 231) km/s. Together with
the three initial masses $M_c(0)$ of clusters and with the three rotational
parameters $\omega_0$ these three positions create the set of 3 $\times$ 3
$\times$ 3, i.e. in total 27 models (see the full list of model parameters in
Table~\ref{model-list_tbl}). Each model comprises a few hundreds of
``snapshots'' made at different moments of time equally spaced from $t=0$ to
$t \gtrsim 1$ Gyr. For each model particle, every snapshot displays its
initial and current mass, coordinates and velocities in the rectangular
Galactocentric coordinate system.

\begin{figure}[]
   \centering
\includegraphics[width=\hsize,clip]{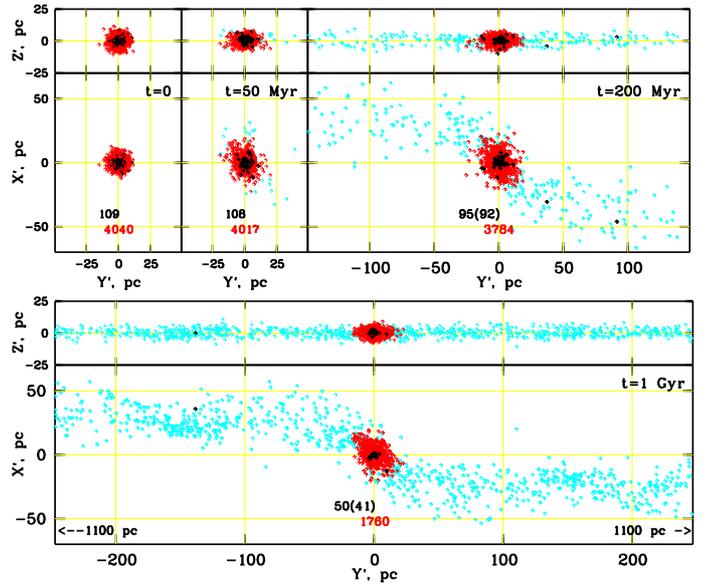}
\caption{The particle distribution in the $X'Y'-$ and $Z'Y'-$ planes of a
model with initial mass $M_c(0) = 10^3 m_\odot$ and distance from the
Galactic centre $R_0$ = 8.5 kpc at different moments of time. The cluster age
is indicated in the right corner of each panel. Cyan dots - all
particles, red dots and red-color labels - dynamically bound particles, black
dots and labels - massive particles ($m> 1 m_\odot$), black labels in
brackets - dynamically bound massive particles. Arrows indicate the limits of
cluster tails when they do not fit to the frame.}
  \label{xyzt-01p_fig}
\end{figure}

To provide the detailed analysis of star cluster shapes we need to find the
cluster  centre position, at first. This issue requires special attention,
since particles having left the cluster during the evolution should not be
taken into account for this purpose. For this we use our own iterative
routine considering the distributions of positions and velocities of
particles and apply it to every snapshot. We find that we are able to
determine the centre even for models with highest mass loss, which lose more
than half of their initial content during the evolution.

As a next step in our analysis, we convert the positions and velocities of
the particles into the proper coordinate system of the cluster related to the
local density centre and define the list of particles which are still bound
to the cluster. We use the particle kinetic and gravitational energy as
criteria of belonging to the cluster. The kinetic energy is computed from the
velocity of every particle in the proper system of the cluster. For the
gravitational energy we use the value of the cluster's self-gravity
potential, determined as the sum of all interactions between a selected
particle and the rest of the particles. We assume, that only the particles
which have a negative relative energy are still bound to the cluster: \[
|E^{gra}_i| > E^{kin}_i. \] In other words, for bound particles we can write:
\begin{equation} |\varphi_i| > \frac{|\mathbf{v}_i|}{2}. \label{dyn-mem_eqn}
\end{equation} Using such a condition for the definition of cluster
membership, we exclude particles which have relatively large velocities
compared to the cluster centre and construct a list of particles which we
select to be ``dynamical'' cluster members.

We define the current mass of the cluster $M_c(t)$ as the sum of masses of
all bound particles. In Fig.~\ref{mass-din_fig} we show the evolution of
$M_c(t)$ for a set of selected models. The evolution shows a similar
behaviour as in the models of \citet{Eetal2007} and \citet{Ketal2008}.
Fig.~\ref{mass-din_fig} indicates, that a star cluster looses from 35\% to
70\% of its initial mass during the first 1 Gyr and the mass loss is
decreasing with increasing distance of clusters from the Galactic centre
and with increasing cluster rotation velocity.

We double check our mass loss sequences by comparing with
the results which we obtain with other fully independent and also
publicly available N-body codes (N-body4 (with GRAPE6) \& N-body6++).
All the models show the very same evolution patterns of mass loss.

\section{Shape parameters of observed and modelled clusters}
\label{obs_mod-res}

\subsection{Shape parameters  of the cluster models}\label{mod_shape_sec}

\begin{figure}[]
   \centering
\resizebox{\hsize}{!}{\includegraphics[angle=270,clip]{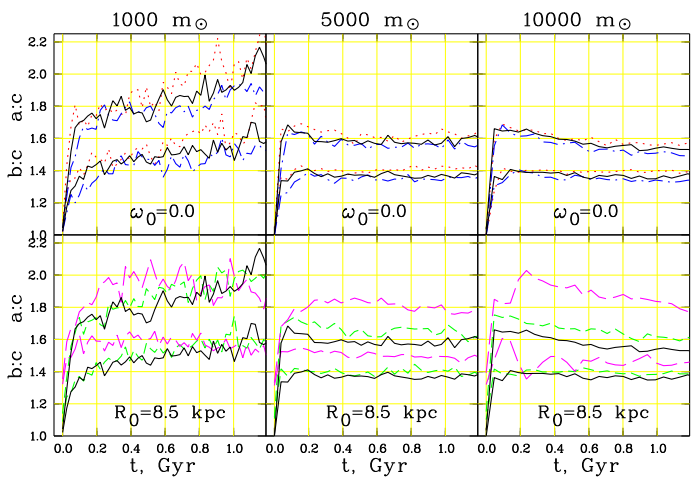}}
\caption{The evolution of ratios of the model ellipsoidal axes $a:c$
and $b:c$ for bound particles for models of different initial mass
(three columns with masses shown at the panel top).
In every panel the upper triplet of the curves indicates the $a:c$ ratio and
the lower triplet corresponds to $b:c$.
The designations are the same as in Fig.~\ref{mass-din_fig}.
}
  \label{abc-din_fig}
\end{figure}

Let us first consider how the cluster model, initialized as a spheroid,
changes its shape and orientation with time. The evolution of a selected
model in the planes $X'Y'$ and $Y'Z'$ is shown in Fig.~\ref{xyzt-01p_fig}.
With time the cluster elongates along the line of the Galactocentric radius,
and begins to ``leak'' losing low-mass particles from the limits
corresponding to the Lagrangian points, farthest and nearest to the Galactic
centre. Because of the differential rotation the particles overtake the
cluster at the end nearest to the Galactic centre and lag it at the farthest
end. The tails contain stars, which are not bound to the cluster anymore. The
dynamical members form an ellipsoid projected onto the $X'Y'$ and $Z'Y'$
planes as ellipses. In the $X'Y'$ plane the ellipse is tilted with respect to
the $X'$-axis at an angle $q_{XY}$, whereas in the $Z'Y'$- plane it follows
the $Y'$-axis.

Such details in the distribution of N-body particles arise due to
interaction with the Galactic tidal field. They are found in all N-body
simulations of open clusters \citep[see, e.g. Fig.7 of ][]{terle87,chura06a}.

In Fig.~\ref{xyzt-01p_fig} one can also clearly see the so-called ``tidal
tail  clumps'' (star density enhancements) at about 150 pc from the cluster
centre. Such clumps were mentioned probably for the  first time in
\citet{capuz05} and were discussed in more detail recently by
\citet{kumah08}. We observe such clumps in all our models (see the model
videos at the FTP link shown in the footnote on page \pageref{models_ref}).
They can be explained by a simple theory \citep{juea08}.

\begin{figure}[]
   \centering
\resizebox{\hsize}{!}{\includegraphics[angle=270,clip]{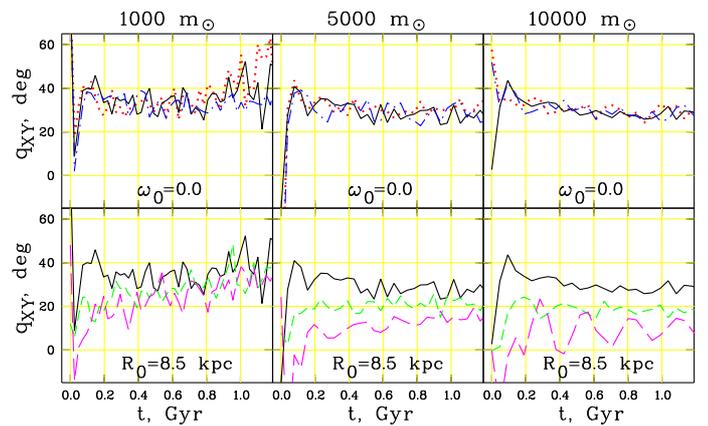}}
\caption{The evolution of the tilt angle $q_{XY}$ computed for dynamically
bound particles for three cases of the initial mass of the cluster models
(three columns with masses shown in the top of the panels). The designations
are the same as in Fig.~\ref{mass-din_fig}.
}
\label{qxy-din_fig}
\end{figure}

In Fig.~\ref{abc-din_fig} and Fig.~\ref{qxy-din_fig} we show the evolution of
the axis ratios $a:c$ and $b:c$ and of the tilt angle $q_{XY}$ for selected
models. Already after a few tens of Myrs the initially spherical cluster
transforms into an ellipsoid tilted with respect to the Galactocentric
radius, with the semi-major axis about twice as large as the minor one, and a
tilt between $30^{\circ}$ and $40^{\circ}$. The shape parameters are almost
independent of the Galactocentric distance of the model clusters, but the
ellipsoid becomes flatter with increasing cluster rotation.

\begin{figure}[]
   \centering
 \resizebox{\hsize}{!}{\includegraphics[angle=270,clip]{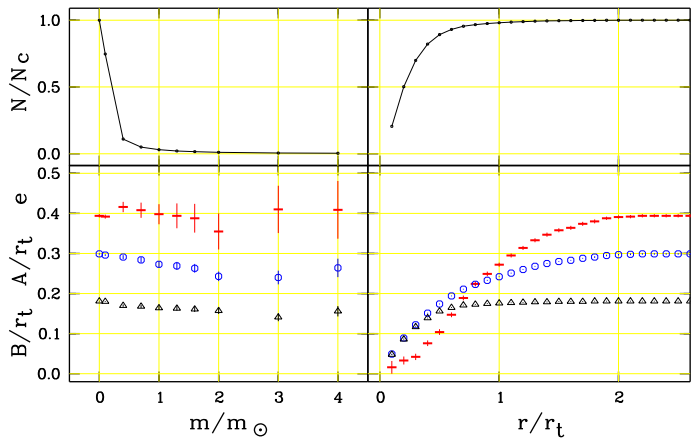}}
\caption{Dependence of the shape of a  model cluster on the sampling
constraints.  Bottom: cluster ellipticity $e$ and semiaxes $A$ and $B$ as
functions of lower mass limit (left) and radius of the considered area
(right). The circles show the major semiaxes, the triangles are for the minor
one, and short (red) horizontal bars mark the ellipticity. Vertical bars
indicate the statistical uncertainties. The upper row shows the numbers of
particles participating in the construction of the corresponding values.}
\label{cl4_fig}
\end{figure}

N-body models offer good possibilities for studying various biases which can
occur in cluster ellipticities derived from observations. In
Fig.~\ref{cl4_fig} we show the dependence of the ellipticity of a model
cluster on sampling constraints due to the mass and the distances from the
cluster centre of the particles considered. In this way we mimic the usual
selection effects which impact the observations of real clusters. Cutting
the model particles at a certain mass limit simulates a possible bias due to
a limited deepness of a survey. The restriction of the models by distances of
particles from the cluster centre simulates the bias  arising due to
underestimation of cluster size. Hereafter, we call the latter effect a
``restricted area bias''. We consider a cluster model, where both major and
minor semi-axes of the ellipsoid are seen under a right angle, i.e. the
projected ellipticity is at its maximum. The model cluster is 150 Myr old,
its current mass is $4.2\cdot10^3\, m_\odot$, and, as a consequence, its
tidal radius is $r_t=22.5$. The number of  dynamically bound particles is
$N_c=19371$.

The tidal radius of the model cluster was computed from the current
total  mass of member particles with the well known relation from
\citet{king62}. This places the model tidal radii into the same system as the
observed tidal radii of open clusters. Note that for about 1/3 of our clusters,
we determined tidal radii by a fit of the cluster radial density distribution to
a King profile \citep{clumart}, and for the remaining clusters we used a proper
calibration to estimate their tidal radii \citep{clumart1}. Then, assuming that
clusters fill up their potential well, the cluster ``tidal'' masses were
computed from the \citet{king62} formula for all 650 clusters.

From Fig.~\ref{cl4_fig} we conclude that excluding low-mass particles from
the consideration has a much smaller impact onto the resulting ellipticity,
than excluding the external parts of the model cluster from consideration.
Indeed,  a removing of low-mass particles causes the axes $A$ and $B$ to
slightly decrease. Since they change, however,  in coordination to each
other, the ellipticity is about constant within the statistical
uncertainties. In contrast, with including  more and more distant particles
in the ellipticity calculation, the ellipticity increases steadily. At a
distance of two tidal radii, the corresponding ellipticity is close to the
limiting value which is achieved if all bound particles within about three
tidal radii are taken into account.

This behaviour can be easily understood if one considers radial dependences
of the axes $A$ and $B$. The minor axis $B$ reaches its maximum value already
within $r_t$ and does not increase anymore, whereas the major axis $A$
continues to rise beyond the tidal radius. One should note that the cluster
flatness is produced by a relatively small number of cluster particles: only
2\% of bound particles occupy a zone beyond $1\,r_t$, and only 0.1\% of them
can be found at $r>2\,r_t$.  In contrast, a few tens of the most massive
particles, which are  not constrained spatially, produce about the same
ellipticity, as the rest of low mass members. A consequence of a smaller
sample is, however, a considerably higher statistical uncertainty in the
resulting ellipticity. A comparison with older models shows that the above
conclusions are still valid for cluster ages up to one Gyr.

In summary we conclude that the determination of cluster ellipticity is not
influenced strongly by the survey's deepness, but rather by the sizes of the
surveyed areas around the clusters. Restricting to areas smaller than the
corresponding tidal radius one takes a risk to introduce a ``restricted area
bias'' in the resulting cluster ellipticities, hence making clusters more
circular than they are in reality.

\subsection{Determination of the uniform shape parameters of the observed and
modelled clusters}

In order to compare the observations with models, one needs parameters
adequately describing both entities. From observations, we compute apparent
ellipticities $e$ and orientation parameters $q$ for each cluster of our
sample using the approach presented in Sec.~\ref{sec_bas-eq} and adopting the
apparent cluster radius $r_2$ from \citet[][]{clucat,newclu}.  The observed
shape parameters $e$ and $q$ for each of the 650 open clusters are listed in
a table that is available in electronic form only. It can be retrieved from
the CDS. To compute the parameters, we consider only the most probable
members, with  kinematic and photometric probabilities higher than 61\%
\citep[see][]{starcat}. This decision was simply guided by the fact that a
sample was stronger contaminated if we included stars with lower membership
probability. Since field stars tend to a random distribution, the
contamination leads to a bias: a fictitious decrease of the  observed
ellipticity.

\begin{figure}[]
   \centering
 \resizebox{0.82\hsize}{!}{\includegraphics[angle=270,clip]{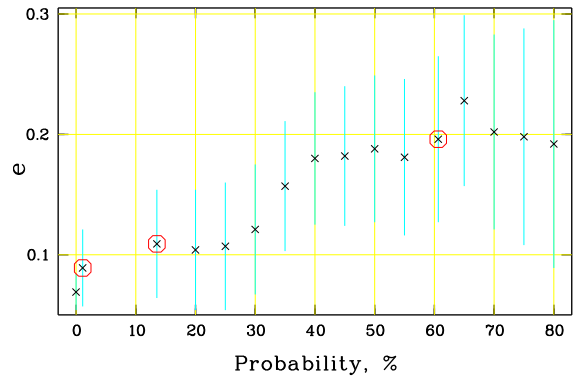}}
\caption{Dependence of the apparent ellipticity $e$ on the probability threshold for
cluster membership for the open cluster Stock~2. The bars correspond to mean
errors. Large open circles indicate the apparent ellipticities based on the
sample of stars with membership probability larger than 1\%(the left
circle), larger than 14\%(the middle circle), larger than 61\%(the right
circle).
}
\label{stock2_fig}
\end{figure}

For illustration, we carried out the ellipticity calculation
considering  cluster stars of different membership probability. For our basic
sample of 152 clusters (see Sec.~\ref{comob_sec} for definition), we obtain an
average ellipticity of $\olin{e}=0.18$ when we consider stars with a membership
probability better than 61\%, $\olin{e}=0.14$ with membership probability better
than 14\%, $\olin{e}=0.12$ with membership probability better than 1\%, and
$\olin{e}=0.09$ if we consider all stars projected on the cluster area. A
probability threshold of 61\% is a compromise between a possible bias due to
backgroud contamination and the number of stars included in the  ellipticity
calculation.

On the other hand, one must make certain that stars with
membership probability larger than 61\% describe adequately the cluster
properties. In Fig.~\ref{stock2_fig} we show the variations of the apparent
ellipticity with the probability threshold in the case of the cluster Stock 2,
for example. This cluster is a well populated cluster, so we get satisfactory
statistics even for the smallest star samples containing the most probable
cluster members. The contamination effect by field stars is strong if the
probability threshold $P$ is less than 30\%, but it becomes almost
insignificant (although still systematic) at $40\% \leqslant P <60\%$. If we
consider stars with membership probabilities larger than 60\%, the variations in
the resulting apparent ellipticity have a random character. This indicates that
- although the sample of stars with membership probability over the 61\%
threshold may contain a number of field stars - they do not introduce a bias in
the apparent ellipticity determination. A further increase of the probability
threshold does not improve the results significantly, but the mean error of the
apparent ellipticities becomes larger due to poor statistics. Therefore, we
consider the choice of a 61\% threshold to be the optimum.

\begin{figure}[]
   \centering
 \resizebox{\hsize}{!}{\includegraphics[angle=270,clip]{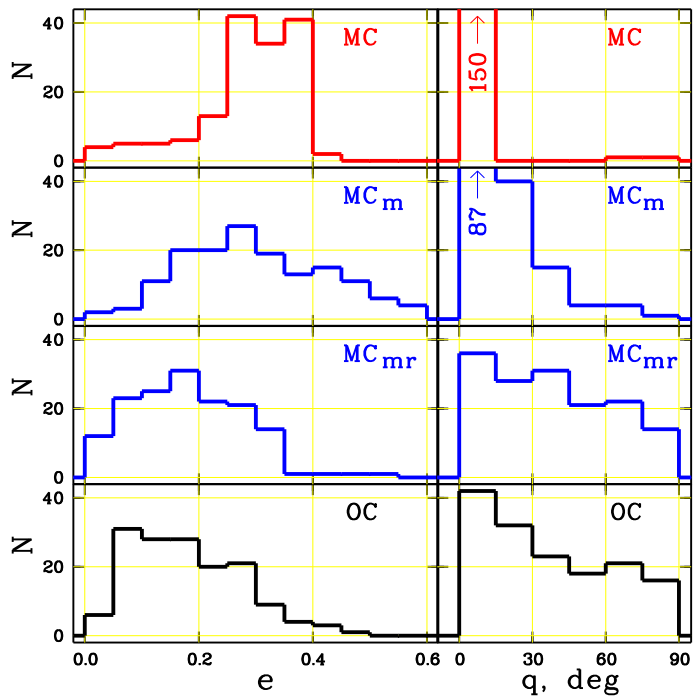}}
\caption{Distributions of ellipticities $e$ (left panels) and orientation
angles $q$ (right panels) for 152 clusters.
The panels are marked with the identifiers of the set.
}
\label{el-q-his_fig}
\end{figure}

From the models, we obtain the axes $a$, $b$, $c$ and the tilt angle
$q_{XY}$ describing the shape and orientation of a cluster in the proper
coordinate system ($X',Y',Z'$). To compare them with the observed parameters
$e$ and $q$, we must ``view'' the models under the same conditions which are
valid for the observed clusters. First, one has to remember that the limiting
magnitude of the \ascc is about $V=12$, and even in the nearest clusters we
can, therefore, observe only stars more massive than $0.7\,M_{\odot}$, i.e.
one order of magnitude higher than the lower mass limit of model particles.
Further, the resulting shape parameters are strongly dependent on the size of
the area around the cluster centre included in the computations. Therefore, a
model cluster must be scaled to the apparent cluster radius $r_2$ of its
observed counterpart.

For each observed cluster, we first selected its theoretical counterpart from
Table~\ref{model-list_tbl} at the Galactocentric distance which best fits the
location of the observed cluster. Then, for a given model we selected a
snapshot with an age closest to the observed age. This snapshot was placed at
the position of the real cluster, and the model particles were projected onto
the face-on plane (i.e. for every particle the standard coordinates $x,y$
were computed). From the complete list of model particles, we selected the
actual members of the cluster by applying eq.~(\ref{dyn-mem_eqn}). We used
this list to compute ``true'' model parameters  of the cluster's shape with
the formulae from Sec.~\ref{sec_bas-eq}.

The models contain  particles down to relatively low masses which can not
always be detected in real observations. In order to reproduce the observed
conditions (i.e. to obtain ``observed'' model parameters), we further
diminished the list of dynamical members selecting particles in the observed
mass range and obeying the condition that the number of modelled and observed
members within the cluster area $r_2$ is approximately the same.

\begin{figure}[]
   \centering
\resizebox{\hsize}{!}{\includegraphics[angle=270,clip]{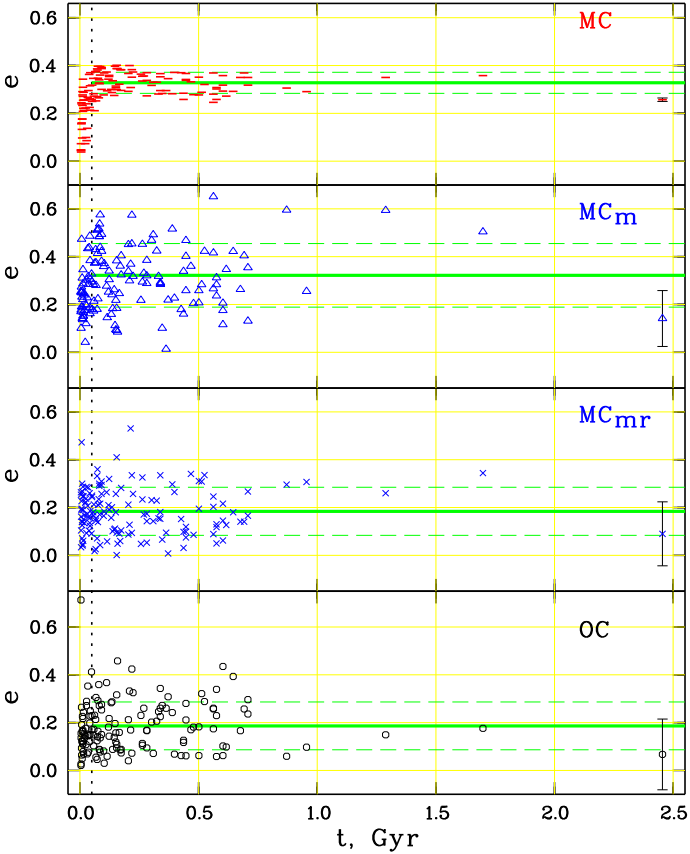}}
\caption{Ellipticities $e$ of 144 low-latitude open clusters ($\beta<20^\circ$)
versus age $t$. The panels are marked with the identifiers of the set.
The dotted vertical line marks $t = 50$ Myr.
In each panel, the horizontal lines show the corresponding average ellipticity
for clusters older than 50~Myr. The dashed lines indicate
their standard deviations.  Vertical bars for a cluster
at $t = 2.46$ Gyr show typical $rms$ errors of the ellipticity estimates.
}
\label{e_age_fig}
\end{figure}

\begin{table*}[t]
\caption{Projected semi-axes $A$ and $B$ in selected directions of the sky
for the model of preferentially oriented cluster ellipsoids.}
\label{axes_tbl}
\setlength{\tabcolsep}{12pt}
\begin{tabular}{ccccc}
\hline\noalign{\smallskip}
Case&$\lambda$&$\beta$&semi-major axis $A$ &semi-minor axis $B$\\
\noalign{\smallskip}
\hline
\noalign{\smallskip}
1)&$0^{\circ}\dots 180^{\circ}$&$ 0^{\circ}$ &$a \cdot b / \sqrt{a^2 \cdot \cos^2 (\lambda-q_{XY})+b^2 \cdot \sin^2 (\lambda-q_{XY})}$&$c$\\
\noalign{\smallskip}
2)&$\lambda_{ac}$&$0^{\circ}\dots 90^{\circ}$&$a$ &$c \cdot b / \sqrt{ b^2 \cdot \cos^2 \beta + c^2 \cdot \sin^2 \beta}$\\
\noalign{\smallskip}
3a)&$\lambda_{bc}$&$0^{\circ}\dots\beta_0$   &$b$&$c \cdot a / \sqrt{a^2 \cdot \cos^2 \beta + c^2 \cdot \sin^2 \beta}$ \\
\noalign{\smallskip}
3b)&$\lambda_{bc}$&$\beta_0\dots 90^{\circ}$ &$c \cdot a / \sqrt{a^2 \cdot \cos^2 \beta + c^2 \cdot \sin^2 \beta}$&$b$  \\
\noalign{\smallskip}
\hline
\end{tabular}
\end{table*}

As a result, we consider four different sets of input data. The first
set includes real observations of clusters. We call this set ``OC''. The
following three sets are related to the models. The set ``MC'' represents
the models as such, and reproduces ``true'' clusters containing all the
dynamical members in the complete range of particle masses. The third set
``$\mathrm{MC_{m}}$'' includes only ``massive'' dynamical members of given
clusters which are above the limiting magnitude of \ascc. Finally, the set
``$\mathrm {MC_{mr}}$'' completely reproduces the observed sample: it
includes the same number of particles selected in the same mass range
$\Delta m$ as we do observe in real clusters. Similar to the observed
clusters, the $\mathrm {MC_{mr}}$ models contain only those particles, which
reside within the apparent cluster radius $r_2$. Since the sets
$\mathrm{MC_{m}}$ and $\mathrm {MC_{mr}}$ bridge the sets MC and OC, they are
introduced to investigate the biases in the observed shape parameters which
may occur due to the limitations of mass ranges and spatial distribution of
cluster stars described in Sec.~\ref{mod_shape_sec}.

\subsection{Comparison of shape parameters of the modelled and observed
clusters}\label{comob_sec}

For comparison with the model data we consider only clusters with more than 20
most probable members. We also exclude the two very populated clusters NGC~869
($h$~Per) and NGC~884 ($\chi$~Per), as they are overlapping in the projection
onto the sky, and have a large number (more than 50\%) of members in common,
making an accurate determination of their individual shape parameters rather
difficult. Hereafter, we refer to the resulting sample as the ``basic sample''
comprising 152 objects.

According to the mass estimates of open clusters given in \citet{clumart1},
the average tidal mass of clusters in the basic sample turns out to be about
$10^3 m_\odot$ which is comparable to the average cluster mass of
$700\,m_\odot$ determined in \citet{fuma} for the complete cluster sample of
650 clusters. Therefore, we  assume that the basic sample  represents
sufficiently well the typical cluster population in the Solar vicinity,
having typical initial masses about $4.5\cdot10^3\,m_\odot$ as determined in
\citet{fuma}. For this reason, we select the cluster models with initial mass
of $5\cdot10^3 m_{\odot}$ as the most suitable ones for comparison with
observations. According to Fig.~\ref{abc-din_fig}, the shape parameters
do not change considerably for models with larger initial masses, whereas the
ellipticity becomes more prominent in clusters with initial masses considerably
lower than the adopted $5\cdot10^3 m_{\odot}$. In the following we consider the
non-rotating models of open clusters keeping in mind that with increasing
rotation velocity clusters become also flatter.

The distributions of ellipticities $e$ and orientation angles $q$ of the
observed and modelled clusters of the basic sample are shown in
Fig.~\ref{el-q-his_fig}. The distribution of ellipticity shows a clear
maximum for the complete cluster  models (the set MC) at $e\approx
0.25\dots0.35$, and the corresponding ellipsoids  are elongated parallel to
the Galactic plane (the orientation angle $q \approx 0^\circ$). For the
models $\mathrm {MC_{m}}$ and $\mathrm {MC_{mr}}$ which take into account the
restrictions set by observations, the ellipticity decreases, while the spread
in $q$ increases indicating a more random orientation of the apparent
ellipses. The observed clusters (the set OC) show relatively small
ellipticities with a peak between $0.075\dots0.175$. Although a peak at $q =
0$  is still observed, the orientation angles $q$ are distributed over the
whole range. This is simply a consequence of small ellipticities since
apparently circular clusters have no orientation angle.

We show the dependence of ellipticities $e$ on cluster age for all
the four sets in Fig.~\ref{e_age_fig}. We limit our consideration to 144
clusters with $\beta<20^\circ$ in order to minimize the impact of the
projection onto the smallest axis of the ellipsoid occurring for clusters at
large galactic latitudes. According to the model set MC, clusters change
their initially circular shapes into ellipsoids during the first 50 Myr and
keep this form thereafter. This behaviour can be also concluded from
Fig.~\ref{xyzt-01p_fig}. The small scatter around the average ellipticity for
clusters older than 50 Myr is defined by their location in the Galactic plane
relative to the Sun and to the Galactic centre, i.e. by the aspect angle
$\lambda$. When we reduce the number of observable members by excluding model
particles of low masses (the set $\mathrm {MC_m}$), the average ellipticity
does not change, although the distribution shows a larger scatter which
arises due to poorer statistics of the remaining dynamical members. When
further excluding model particles located outside the apparent radius assumed
for the clusters, the average ellipticity becomes significantly
smaller. The observed clusters show a similar ellipticity distribution as the
model set $\mathrm{MC_{mr}}$. These features confirm the analysis carried out
in Sec.~\ref{mod_shape_sec}.

The following statistics supports the decisive role of spatial limitation ($r
\le r_2$) hiding the real ellipticity of clusters. The average ellipticities of
103 clusters with $t > 50$ Myr are equal to $0.328 \pm 0.004$, $0.322 \pm
0.013$, $0.184 \pm 0.010$, $0.186 \pm 0.010$ for MC, $\mathrm {MC_{m}}$,
$\mathrm {MC_{mr}}$ and CO, respectively. The similar average
ellipticities of the sets $\mathrm {MC_{mr}}$ and OC indicate that a possible
bias due to background contamination is rather small if the ellipticity
calculation is based on cluster stars with membership probabilities higher than
61\%. We conclude, also, that a spatial limitation of cluster members
introduces a significant bias in the determination of ellipticities, whereas a
mass limitation increases mainly the random uncertainties of the results. The
smaller ellipticities of the sets $\mathrm {MC_{mr}}$ and OC  are results of too
small radii $r_2$ assumed for the clusters. Note that, on average, the
\ascc-based estimates of cluster radii $r_2$ \citep{clucat} are already larger
by a factor of two compared to previously published values. Better estimates of
cluster radii are only possible if a proper separation of cluster members from
the numerous field stars can be achieved in outer cluster regions. This would
require more accurate surveys of proper motions and photometric data all over
the sky. On the other hand, surveys deeper than the \ascc would increase the
random accuracy of the ellipticity determination of open clusters.

\subsection{Distribution of the cluster ellipticity over the sky}

Up to now, the ellipticity of open clusters was studied either for
individual  objects \citep{ramer98a,ramer98b}, or in selected directions of
the  Galactic disk \citep{chen04}. In the present study we determined the
apparent shape parameters for clusters observed all over the sky. Therefore,
it seems to be useful to derive geometric  relations between the apparent
ellipticity of clusters and their spatial location in the Galaxy. Of course,
a regular dependence can only be expected if cluster ellipsoids are not
orientated randomly in space but show a certain orientation with respect to
the Galactic centre.

Indeed, the model calculations indicate that the Galactic tidal field and
differential rotation  quickly align clusters along a preferential direction,
and the resulting ellipsoids show always the same axes ratios. Let
$a,\,b,\,c$ be the semi-axes of these three-axial ellipsoids with the
semi-major axis $a$ tilted with respect to the Galactocentric radii of the
clusters by an angle $q_{XY}$. For simplicity, we further assume that the
major axis is parallel to the  Galactic plane. Hereafter, we call this
approach the  model of  preferentially-oriented ellipsoids, and the
ellipsoid itself the ``reference'' one.

From observations, however, we can only see a two-dimensional projection of
the reference ellipsoid on the sky that is an ellipse with the semi-axes $A$
and $B$. The ratio $B/A$  varies from $c/a$ to 1, depending on the aspect
angles $\lambda$ and $\beta$ describing the cluster location with respect to
the Sun and the Galactic centre. The apparent ellipticity $e$ changes from 0
to $1 - c/a$, respectively.

\begin{figure}[]
   \centering
\resizebox{\hsize}{!}{\includegraphics[angle=270,clip]{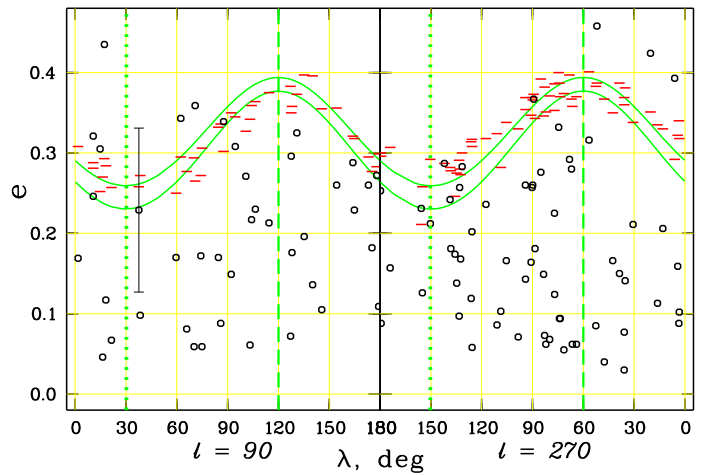}}
\caption{Ellipticity $e$ versus aspect angle $\lambda$ for 103 clusters with
$\beta < 20^{\circ}$ and $t > 50$~Myr. The left panel corresponds to the
Galactic longitude range $l = 0^{\circ}...180^{\circ}$, the right panel to $l
= 180^{\circ}...360^{\circ}$. The curves show the
$e(\lambda,\beta)$--relations for $\beta=0^\circ$ (upper curve) and
$\beta=20^\circ$ (lower curve) constructed for the reference ellipsoid tilted
to the direction to the Galactic centre by angle $q_{XY} = 30^\circ$, having
axes ratios (1.65 : 1.35 : 1). Short horizontal bars (red): set MC (complete
models), open circles: set OC (observations). The vertical bars drawn for one
object show the typical
$rms$ error of the observed ellipticity. The dashed vertical lines mark
$\lambda_{ac}$, the dotted ones $\lambda_{bc}$. }
\label{e-lambda_fig}
\end{figure}

When a cluster is located in the Galactic plane ($b = \beta = 0^{\circ}$), $B
= c$ is always valid, and $A$ varies from $b$ to $a$. The apparent
ellipticity reaches the maximum $e = 1 - c/a$ at aspect angles
$\lambda_{ac}=90^\circ+q_{XY}$ (for $l=0^\circ\dots180^\circ$) and at
$\lambda_{ac}=90^\circ-q_{XY}$ (for $l=180^\circ\dots360^\circ$). In
contrary, the apparent ellipticity reaches the minimum $e = 1 - c/b$ at
$\lambda_{bc}=q_{XY}$ (for $l=0^\circ\dots180^\circ$) and
$\lambda_{bc}=180^\circ- q_{XY}$  (for $l=180^\circ\dots360^\circ$). For all
other directions, the apparent ellipticities can be computed from the
equations given in the first row of Table~\ref{axes_tbl}.

Though the majority of star clusters is located near the Galactic plane,
there is a number of clusters at higher galactic latitudes.
In a special case, when clusters are
located directly at the Galactic Poles, one observes $A = a$ and $B = b$. For
all other clusters outside  the Galactic plane, the projection of the
reference ellipsoid is defined by both aspect angles. For example, at
$\lambda=\lambda_{ac}$ the ellipticity decreases monotonically from $1-c/a$
at $\beta=0^\circ$ to $1-b/a$ at $\beta=90^\circ$. The corresponding
expression is given in the second row of Table~\ref{axes_tbl}. At
$\lambda=\lambda_{bc}$ this relation is rather different since the projected
ellipse changes its orientation at some angle $\beta_0$. Thus the
ellipticity first decreases from $1-c/b$ at $\beta=0^\circ$ to 0 at
$\beta=\beta_0$, and then increases to $1-b/a$ at $\beta=90^\circ$. The
corresponding equations are listed in the last two rows of
Table~\ref{axes_tbl}. The aspect angle $\beta_0$ where the projection becomes
a circle ($A = B$) is determined by the semi-axes $a$, $b$, $c$ of the
reference ellipsoid from the equality condition

\begin{equation}
b = \frac{c \cdot a}
{\sqrt{a^2 \cdot \cos^2 \beta_0 + c^2 \cdot \sin^2 \beta_0}}. \label{betac_eq}
\end{equation}

For given parameters of the reference ellipsoid, the Galactic meridian at
$\lambda_{ac}$ defines a locus of maximum ellipticities
$e(\lambda_{ac},\beta)$ over the sky, whereas the Galactic meridian at
$\lambda_{bc}$ gives a locus of minimum ellipticities
$e(\lambda_{bc},\beta)$.

\begin{figure}[]
   \centering
\resizebox{\hsize}{!}{\includegraphics[angle=270,clip]{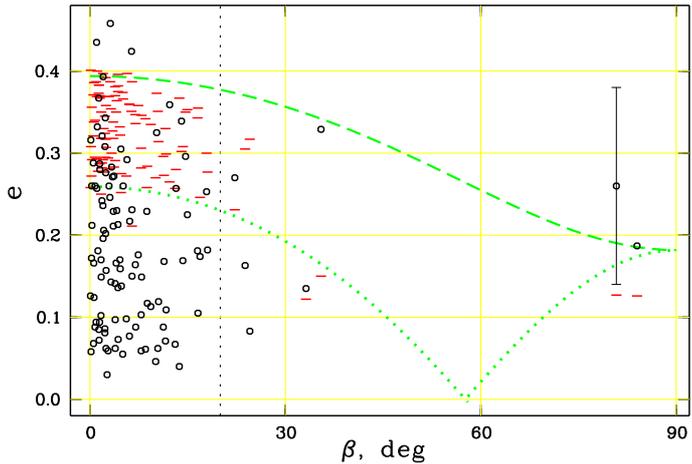}}
\caption{Ellipticity $e$ versus aspect angle $\beta$ for 110 clusters with
$t>50$ Myr. The curves show the $e(\lambda,\beta)$--relations for
$\lambda=\lambda_{ac}$ (dashed curve) and $\lambda=\lambda_{bc}$ (dotted curve)
constructed for the reference ellipsoid tilted to the direction of the Galactic centre by
an angle $q_{XY} = 30^\circ$, and with axes ratios (1.65 : 1.35 : 1). Short
horizontal bars (red):  MC (complete models), open circles:  OC (observations).
The vertical bars drawn for one
object show the typical $rms$
error of the observed ellipticity.
The dashed vertical line marks $\beta=20^\circ$.
}\label{e-beta_fig}
\end{figure}

In Fig.~\ref{e-lambda_fig} and Fig.~\ref{e-beta_fig} we plot the analytical
relations  from Table~\ref{axes_tbl} together with the model and observed
data on ellipticities versus the aspect  angles for ``relaxed'' clusters
which are older than 50 Myr. Though a few observed ellipticities fit the
predicted values, the majority of the observed clusters show, as expected,
too low ellipticities due to the restricted area bias described above.
Therefore, we cannot use the observations for deriving the parameters of
the reference ellipsoid. Instead, the parameters $a,\,b,\,c$ and $q_{XY}$
were found by fitting the relations from Table~\ref{axes_tbl} to the data
points of the complete models (MC) as $(a:b:c) =(1.65:1.35:1)$ and $q_{XY} =
30^{\circ}$. These parameters are based on the non-rotating models assuming
the same initial masses ($5\cdot10^3 M_{\odot}$) but different location of
clusters from the Galactic centre (see also Fig.~\ref{abc-din_fig} and
Fig.~\ref{qxy-din_fig}). For this reference ellipsoid, the angle $\beta_0$ is
determined from eq.~(\ref{betac_eq}) to be $\beta_0 = 57.6^{\circ}$.

\subsection{Comparison with the literature}

\begin{figure}[]
   \centering
\resizebox{\hsize}{!}{\includegraphics[angle=270,clip]{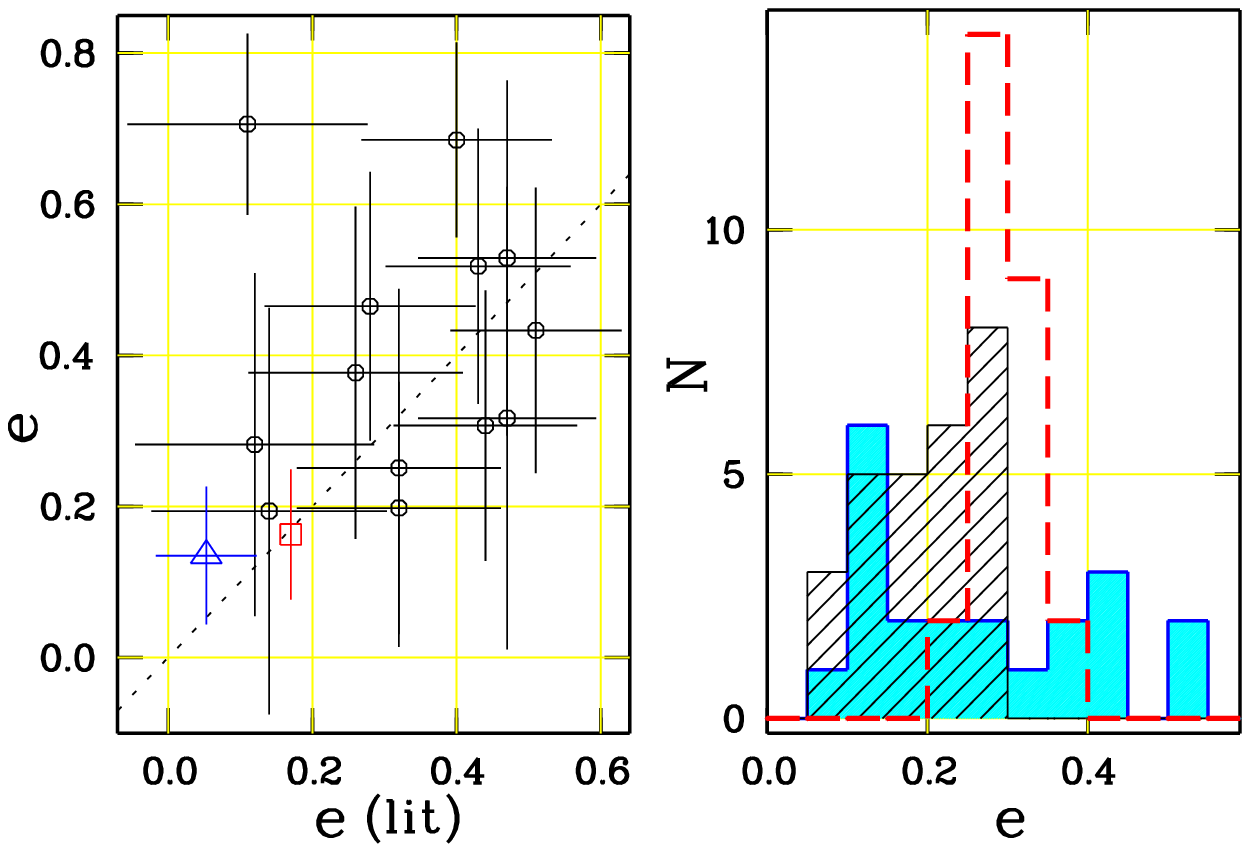}}
\caption{Comparison of our results with data from the literature. The left panel
compares ellipticities of selected clusters. The clusters from \citet{chen04}
are shown with open circles, the Pleiades \citep{ramer98a} with a red square,
and Praesepe \citep{ramer98b} with a blue triangle. The dotted line is the
bisector. Right panel: comparison of distributions with $e$ for 21 clusters
with $t>50$ Myr within the Galactic longitude range
$144^{\circ}...250^{\circ}$ of \citet{chen04} -- filled histogram, and our 27
clusters within the same longitude and age ranges (hatched histogram).
The dashed (red) histogram is the distribution of full models (set MC)
of 27 clusters. See details in the text.}
\label{compbi_fig}
\end{figure}

In the literature, there are only a few studies on the ellipticity of
Galactic open clusters. \citet{ramer98a} and \citet{ramer98b} studied the
shape of the Pleiades and Praesepe using the best data on astrometric and
photometric membership then available for these open clusters. For the
Pleiades, \citet{ramer98a} determined $e = 0.17\pm0.05$. Since
\citet{ramer98b} do not provide the ellipticity for Praesepe, we computed
this value ($e=0.05\pm0.07$) based on the data on the cluster membership from
their Table~2. \citet{chen04} published morphological parameters, including
the ellipticity, of 31 Galactic open clusters, residing basically in the
Galactic anticentre direction. They used the 2MASS catalogue and segregated
clusters from the background with the help of an equidensity method. We
compare our results with the above studies in Fig.~\ref{compbi_fig}.

Our results for the two nearest clusters ($e=0.16\pm0.09$ for the Pleiades,
and $e=0.14\pm0.09$ for Praesepe) are in good agreement with the
ellipticities derived by \citet{ramer98a} and \citet{ramer98b}. This can be
expected since both studies are based on the catalogues of the
Hipparcos-Tycho family, and the member lists of Raboud and Mermilliod
practically coincide with our membership for these clusters. Because the
Pleiades and Praesepe are at small distances from the Sun and show large
proper motions, one can reliably determine membership in these clusters up to
relatively large distances from the cluster centres and down to stars of
relatively low masses. Therefore, these clusters are the best candidates to
get realistic ellipticities from observations. Indeed, the complete models
(MC) provide $e=0.30$ and 0.12 for the Pleiades and Praesepe, respectively,
and fit the observations reasonably well.

A comparison with the results of \citet{chen04} is expected to be more
informative since different techniques and different input data are used for
ellipticity determination. Unfortunately, both studies have only 13 clusters
in common, and 12 of them do not belong to the basic sample since the number
of their most probable members is less than 20 stars in our data.
Nevertheless, the agreement between the both results is reasonable: the
ellipticities of 11 of 13 clusters deviate from the bisector by less than one
$rms$-error in Fig.~\ref{compbi_fig} (left panel).

In Fig.~\ref{compbi_fig} (right panel) we compare the distributions of
cluster ellipticities of \citet{chen04} with our models (MC) and observations
(OC). In order to avoid coordinate-dependent mis-interpretations, we consider
only those clusters from our basic sample that are located in the same area
of the sky as the clusters of the sample of \citet{chen04}, i.e.
$l=144^\circ\dots250^\circ,\,b=-18^\circ\dots20^\circ$.
We also exclude all clusters younger than 50 Myr from the comparison.
The restricted samples
include 21 clusters of \citet{chen04} and 27 of our clusters with observed
and model ellipticities. Whereas N-body models forecast an average
ellipticity of $e\approx 0.3$ and a narrow spread of ellipticities in this
Galactic direction, our observations show a much wider distribution enhanced at
smaller ellipticities ($e = 0.06...0.29$). Provided that the peak in the
distribution of \citet{chen04} data at low ellipticities ($e\approx
0.10-0.15$) is not random, we conclude that their data suffer even more from the
restricted area bias than our observations do. The large spread in the data of
\citet{chen04} can be explained by larger random errors of their
ellipticities.

We briefly mention here results on the ellipticities of Galactic globular
clusters. For the latter \citet{white87} determined shape parameters based
on  the equidensity contours method from the data of Palomar and SRC Sky
surveys. They find that for 99 globular cluster the average ellipticity is
only $0.07\pm 0.01$, and the orientation angles are distributed randomly
(because the clusters are almost round). \citet{white87} point  to
interstellar extinction as a probable reason of such an unexpected  result.

\section{Conclusion}
\label{sec_concl}

Based on a multicomponent analysis of coordinates of the most probable
cluster members we have determined shape parameters of 650 Galactic open
clusters. Ellipticities and orientation angles complete the list of
morphological and dynamical parameters (core sizes and apparent cluster
radii, indicators of mass segregation, tidal radii) which were determined and
analysed in a series of previous papers \citep[][]{clusim,clumart,clumart1}.

We have carried out high resolution N-body simulations with the specially
developed $\phi$GRAPE code on the parallel GRAPE systems developed at
ARI Heidelberg and MAO Kiev. The set of 27 cluster models (3 initial masses
$\times$ 3 Galactocentric distances $\times$ 3 rotation parameters) with a
maximum number of particles of 40404 was evolved during 1 Gyr. The cluster
particles are initially distributed according to a Salpeter IMF in the mass
range $0.08...8.0~m_\odot$.  For each particle, the decision on its
cluster membership is made by comparing its potential and kinetic energies.

The calculations of the apparent shape parameters of the modelled and
observed clusters were carried out with the same technique. The selection of
suitable models was based on initial cluster masses that were estimated from
the earlier studies of our cluster sample \citep[][]{clumart,clumart1}. This
enabled us to realize a self-consistent approach for comparison of model and
observed shape parameters using the data on 152 of the most populated
Galactic clusters from our sample.

N-body calculations show that all the models loose more than about 50\% of
their initial mass during the first Gyr of the evolution due to two body
encounters only. A model cluster older than $\approx 50$~Myr keeps an oblate
shape, with the major axis $a$ tilted with respect to the Galactocentric
radius at an angle of $\approx$30-40 degrees. The model counterparts
corresponding to 152 observed clusters have an ellipticity distribution with a
peak at $e \approx 0.3..0.4$, and, on average, the ellipticity becomes
even larger for rotating model clusters.

However, the observed clusters show a significantly lower ellipticity,
typically of about $e=0.2$. A comparison of observed cluster shapes with the
detailed models allowed us to realize that lower ellipticities are only
apparent. We explain this disagreement  by a bias in observations due to
underestimated cluster sizes $r_2$ based on the \ascc data and used in the
determination of cluster ellipticities. According to the models, the
ellipticity of the central region of a cluster is small, it increases if
outer layers are taken into account, and approaches asymptotically the
largest value at 2-3 tidal radii $r_t$. Particles outside the tidal radius
contribute negligibly to the cluster mass, however, their spatial
distribution is decisive for the determination of the cluster ellipticity.
Due to a relatively low density of cluster members above the background  in
outer cluster regions, real clusters are usually observable only within
distances smaller than $r_t$ from the cluster centres.  Though the apparent
cluster sizes $r_2$ based on the \ascc are twice as large as  previously
published in the literature, the ratio $r_2/r_t$ was found to be, on average,
only 0.5 \citep{clumart1}.

In contrast, the deepness of the survey does not play such a dramatic role
for the ellipticity determination. Although  deeper surveys offer a
possibility to identify  more numerous cluster members of lower masses and,
consequently, to increase the random accuracy of the ellipticity
determination, this alone is not sufficient to avoid  the ``restricted area
bias''.

We adopt the assumption of \citet{wiel74,wiel85} on the preferential shape
and orientation of cluster ellipsoids in the Galaxy and derive simple
formulae giving a relation between the apparent ellipticity and spatial
location of clusters of our sample. Due to relatively high systematic and
random errors in observed ellipticities, however, we could not use the
present observations for estimating the parameters of the reference
ellipsoid. According to N-body calculations, the cluster ellipsoid has an
axes ratio of $a:b:c=1.65:1.35:1$, and it is tilted by an angle $q_{XY}
\approx 30^\circ$ with respect to the Galactocentric radius.

In order to test the models with the real observations, the ellipticity
determination for clusters must be based on sufficiently large areas (up to
two to three tidal radii) around the clusters. From this point of view,
equidensity methods are hardly useful to determine real cluster sizes,
especially, due to a low density contrast of cluster members above the
background at large distances from the cluster centres. The membership
determination with kinematic and photometric criteria fits the problem better
but requires surveys with higher accuracy of proper motions than it is in the
\ascc.

\begin{acknowledgements}

We would like to thank Rob Jeffries, the referee, for his useful
comments and suggestions, which helped us to improve the paper.

This study was supported by DFG grant 436 RUS 113/757/0-2, and RFBR grants
06-02-16379, 07-02-91566.

P. Berczik \& M. Petrov thank for the special support of his work by the
Ukrainian National Academy of Sciences under the  Main Astronomical
Observatory GRAPE/GRID computing cluster  project.

P. Berczik acknowledges his support from the  German Science Foundation (DGF)
under SFB 439 (sub-project B11) \textsl{``Galaxies in the Young
Universe''} at the University of Heidelberg. His work was also supported  by
the Volkswagen Foundation \textsl{``Volkswagenstiftung''} under GRACE
Project and grant No. I80 041-043 of the  Ministry of Science, Education and
Arts of the state of Baden-W\"urttemberg, Germany.

\end{acknowledgements}

\bibliographystyle{aa}
\bibliography{clubib}

\end{document}